\documentclass[conference]{IEEEtran}
\IEEEoverridecommandlockouts
\usepackage{cite}
\usepackage{amsmath,amssymb,amsfonts}
\usepackage{algorithmic}
\usepackage{graphicx}
\usepackage{textcomp}
\usepackage{xcolor}
\usepackage{booktabs,tabularx}
\usepackage{tcolorbox}
\usepackage{tikz}
\usetikzlibrary{positioning,arrows.meta,fit,backgrounds}
\usepackage[hidelinks]{hyperref}

\def\BibTeX{{\rm B\kern-.05em{\sc i\kern-.025em b}\kern-.08em
    T\kern-.1667em\lower.7ex\hbox{E}\kern-.125emX}}

\makeatletter
\newcommand{\linebreakand}{%
  \end{@IEEEauthorhalign}
  \hfill\mbox{}\par
  \mbox{}\hfill\begin{@IEEEauthorhalign}
}
\makeatother

\begin{document}
\bstctlcite{IEEEexample:BSTcontrol}

\title{Reusable LLM-Based Plans with Probabilistic Guarantees for Self-Adaptive Systems}

\author{
\IEEEauthorblockN{Asher Tempest Arya}
\IEEEauthorblockA{\textit{University of Toronto}\\
Toronto, Canada\\
asherarya@cs.toronto.edu}
\and
\IEEEauthorblockN{Rajan Maghera}
\IEEEauthorblockA{\textit{University of Toronto}\\
Toronto, Canada\\
maghera@cs.toronto.edu}
\and
\IEEEauthorblockN{Nicholas Wu Jiang}
\IEEEauthorblockA{\textit{University of Toronto}\\
Toronto, Canada\\
nwjiang@cs.toronto.edu}
\linebreakand
\IEEEauthorblockN{Federico Mora}
\IEEEauthorblockA{\textit{University of Waterloo}\\
Waterloo, Canada\\
federico@uwaterloo.ca}
\and
\IEEEauthorblockN{Aren A. Babikian}
\IEEEauthorblockA{\textit{McGill University}\\
Montreal, Canada\\
aren.babikian@mcgill.ca}
\and
\IEEEauthorblockN{Marsha Chechik}
\IEEEauthorblockA{\textit{University of Toronto}\\
Toronto, Canada\\
chechik@cs.toronto.edu}
}

\maketitle

\begin{abstract}
Self-adaptive systems are designed to accommodate variants of their environment, which form a family of instances whose uncertainty is captured by probabilistic models, and must meet probabilistic requirements on each instance.
Within such a family, changes in size and structure complicate the fundamental adaptation decision: whether to retain, reuse, or construct a new controller, since controllers are often synthesized for and specific to an instance.
Large language models (LLMs) have recently been used to write reusable controllers, but existing approaches largely target deterministic domains, and probabilistic model checking of their output remains limited to preliminary single-instance studies.
In this paper, we propose a verifier-guided framework for controller reuse with explicit support for families of environments.
The framework introduces a novel combination of LLM-written controllers with probabilistic model checking to provide per-instance guarantees for probabilistic requirements.
We evaluate our approach using three LLMs on three benchmarks, assessing its effectiveness and cost with respect to baseline implementations, the distribution of adaptation decisions, and the impact of individual framework components.
On grid worlds, LLM-written rule lists are verified on about 90\% of held-out instances, against at most 77\% for the baselines, and verifying a stored rule list is 15 to 103 times faster than per-instance synthesis.
Reuse serves most episodes at a small fraction of the cost of generation or per-instance synthesis, and model-checking feedback increases verified coverage over independent restarts.
\end{abstract}


\section{Introduction}
\label{sec:intro}
\textbf{Motivation.}
Self-adaptive systems adjust their behavior at run time to keep meeting their requirements as their environment changes, typically through a MAPE-K loop~\cite{Kephart2003-MAPE-K}.

A central adaptation decision is whether to retain the current controller, reuse another, or construct a new one.
The decision is hard when the environment changes in size and structure, e.g., for a robot deployed in buildings whose maps differ, or for a Mars rover whose resource arbiter serves different numbers of requests. 
We call each such environment, e.g., one building, an \textit{instance} of a \textit{family} of environments. 
When elements inside the environment are uncertain, e.g., a robot's {actions} can {fail} or other agents act unpredictably, probabilistic requirements, such as the upper bound on the probability of a collision, are required.

\textbf{Problem Statement.}
Before each run on an instance, the system must deploy a controller that is guaranteed to meet the probabilistic requirements on the model of that instance, which {may be completely new.}
Probabilistic model checking at run time can establish whether a configuration of the system meets such requirements~\cite{Calinescu2012-CACM-runtime-QV,Calinescu2018-ENTRUST}.
Synthesizing a strategy for each instance with a probabilistic model checker such as PRISM~\cite{Kwiatkowska2011-PRISM} or Storm~\cite{Hensel2022-Storm} gives this guarantee, but at the cost of reuse and speed.
Each strategy applies only to its own instance.
Synthesizing the full model of each new instance is slow, and delays every adaptation.
Methods for families of Markov decision processes (MDPs) synthesize strategies that meet the requirements on every member~\cite{Andriushchenko2024-PoliciesGrowOnTrees,Heck2026-SMPMC}, but assume that all members share one state space.

Existing reusable controllers avoid repeated synthesis, but they are limited in their guarantees or in how far they generalize~\cite{Fern2006-decision-lists,Cobbe2020-Procgen,Azeem2025-123Go}.
Large language models (LLMs) can write reusable controllers as generalized plans~\cite{Silver2024-GenPlan-LLM,Stein2026-strategy-refinement,Drexler2026-PromptProvePatch,Murray2026-GenePlan}, but these generalized planners target deterministic domains, in which simply running a plan on an instance verifies it.
Where probabilistic model checking has been applied to LLM output, it covers a single model~\cite{Gross2025-verifying-LLM-decisions,Lee2025-VeriPlan,Gross2024-counterfactual-LLM,Arya2026-gridworld-PMC}.

To efficiently and verifiably deploy adaptive LLM-written controllers in uncertain environments, architectures must integrate LLM reasoning with formal probabilistic verification, supporting temporal and probabilistic requirements.

\textbf{Contributions.}
In this paper, we propose a \textit{verifier-guided framework for adaptation through controller reuse and generation} across families of stochastic environments. Our specific contributions are as follows:
(1) We propose a \emph{verifier-guided} adaptation mechanism that retains the current controller only if it is verified on the instance at hand, and otherwise selects a verified one from a library or generates a new one and adds it to the library.
(2) We define \textit{generalized rule lists}, a formal representation of reusable symbolic controllers (\autoref{fig:intro-rules}), and a permissive strategy criterion ~\cite{Draeger2015-permissive} for verifying them on an instance.
(3) We propose an \emph{LLM-based approach} for generating and repairing controllers with \emph{verification feedback.}
(4) {We empirically assess verified coverage and the effect of verifier feedback, and compare the cost of reuse with that of synthesis per instance on grid-world navigation, Pac-Man~\cite{Hartmanns2019-QVBS} and Mars Exploration Rovers (MER)~\cite{Feng2011-MER-model}.}
We compare LLM-written rule lists with decision trees learned as in 1--2--3--Go!~\cite{Azeem2025-123Go} and with policies trained by reinforcement learning.

\begin{figure}[t]
\centering
\footnotesize
\setlength{\tabcolsep}{2pt}
\begin{tabularx}{\columnwidth}{@{}r>{\raggedright\arraybackslash}Xl@{}}
\toprule
\# & Condition & Action \\
\midrule
1 & \texttt{g3} & \texttt{any} \\
2 & \texttt{!wall \& closer \& !risky \& !obstacle \& !early} & \texttt{any} \\
3 & \texttt{!wall \& !obstacle \& !early \& !risky \& !patrol} & \texttt{any} \\
4 & \texttt{!wall \& closer \& !obstacle \& !early \& !patrol} & \texttt{any} \\
5 & \texttt{!wall \& !obstacle \& !early \& !risky} & \texttt{any} \\
6 & \texttt{wall \& !obstacle \& !early \& !risky} & \texttt{any} \\
7 & \texttt{!wall \& !obstacle \& !early} & \texttt{any} \\
8 & \texttt{true} & \texttt{any} \\
\bottomrule
\end{tabularx}
\caption{A rule list written by Claude Haiku~5.5 for three members (6$\times$6 to 8$\times$8) of a rooms family of grid worlds.}
\label{fig:intro-rules}
\vspace{-0.15in}
\end{figure}

\textbf{Novelty and Significance.}
{To the best of our knowledge, \textbf{this paper is the first to combine LLM-written controllers with \emph{probabilistic verification} to adapt across \emph{stochastic environments} under \emph{temporal logic constraints}.}}
Our results provide insight into how far these controllers generalize across such environments and the costs of such adaptation decisions. On grid worlds, LLM-written rule lists are verified on about 90\% of held-out instances, against at most 77\% for the baselines, and verifying a stored rule list is 15 to 103 times faster than per-instance synthesis.
Furthermore, since each guarantee comes from verification, our framework lets self-adaptive systems use LLM-written controllers without assuming that the LLM's output meets the requirements.
\section{Background}
\label{sec:background}
\subsection{Markov Decision Processes}
\label{sec:bg-mdp}

An MDP is a tuple $M = (S, s_0, A, P)$, where $S$ is a finite set of states, $s_0 \in S$ is the initial state, $A$ is a finite set of actions, and $P : S \times A \times S \to [0,1]$ gives the probability $P(s, a, s')$ of moving from $s$ to $s'$ when action $a$ is taken.
An action $a$ is \textit{available} in a state $s$ if $P(s, a, \cdot)$ is a probability distribution, and otherwise $P(s, a, s') = 0$ for every $s' \in S$.
We write $A(s) \subseteq A$ for the non-empty set of actions available in $s$.
A \textit{strategy} $\sigma$ resolves the choice among the available actions in each step, may use memory and randomization, and turns $M$ into the Markov chain that $\sigma$ induces~\cite{Baier2008-principles}.

A \textit{parameterized MDP} describes a set of MDPs, one for each \textit{instance}, i.e., each value of its parameters, such as the size of a map or the length of a game~\cite{Azeem2025-123Go}.

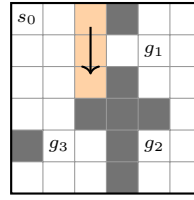
\begin{figure}[t]
\centering
\vskip 0.1in
\begin{minipage}[c]{0.3\columnwidth}
\centering
\begin{tikzpicture}[scale=0.42, every node/.style={font=\scriptsize}]
\foreach \c/\r in {3/0, 3/2, 2/3, 3/3, 4/3, 0/4, 3/4, 3/5}
  \fill[black!55] (\c, -\r) rectangle ++(1, -1);
\foreach \r in {0, 1, 2}
  \fill[orange!35] (2, -\r) rectangle ++(1, -1);
\draw[step=1, black!40] (0, 0) grid (6, -6);
\draw[thick] (0, 0) rectangle (6, -6);
\node at (0.5, -0.5) {$s_0$};
\node at (4.5, -1.5) {$g_1$};
\node at (4.5, -4.5) {$g_2$};
\node at (1.5, -4.5) {$g_3$};
\draw[->, thick] (2.5, -0.75) -- (2.5, -2.25);
\end{tikzpicture}
\end{minipage}\hfill
\begin{minipage}[c]{0.67\columnwidth}
\caption{A 6$\times$6 instance of the grid world. The agent starts in cell $s_0$ and must visit $g_1$, $g_2$ and $g_3$ in order. Dark cells are obstacles, and a moving obstacle patrols the shaded cells, down and back up.}
\label{fig:grid}
\end{minipage}
\vspace{-0.15in}
\end{figure}

\textbf{Running Example.}
We use the stochastic grid world of prior work~\cite{Arya2026-gridworld-PMC}, a parameterized MDP whose parameters are the grid's size and layout (\autoref{fig:grid}).
An agent starts in the top-left cell of an $n \times n$ grid and must visit the goals $g_1, g_2, g_3$ in order.
In each step, it chooses to move up, right, down or left.
A move succeeds with probability 0.7 and slips to either side with probability 0.15 each, and a move off the grid or into a static obstacle leaves the agent in place.
A moving obstacle patrols a fixed cycle of cells, advancing one cell per step.
A state records the agent's cell, which goals it has reached (the \textit{goal flags}), and the moving obstacle's position.

\subsection{Probabilistic Model Checking}
\label{sec:bg-pmc}

Probabilistic model checkers such as PRISM~\cite{Kwiatkowska2011-PRISM} compute the probability $\mathit{Pr}^{\sigma}(\psi)$ of the paths from $s_0$ that satisfy a temporal property $\psi$ in the Markov chain that a strategy $\sigma$ induces.
We express properties as path formulas of linear temporal logic~\cite{Pnueli1977-LTL} over atomic propositions that label states, using the operators \textit{eventually} ($\mathbf{F}$), \textit{always} ($\mathbf{G}$) and \textit{until} ($\mathbf{U}$).
A \textit{requirement} $\varphi = P_{\bowtie \theta}[\psi]$, with $\bowtie\, \in \{\geq, \leq\}$, bounds the probability of $\psi$ by a threshold $\theta$.
A strategy $\sigma$ \textit{satisfies} $\varphi$ if $\mathit{Pr}^{\sigma}(\psi) \bowtie \theta$.
The \textit{worst case} of a requirement on an MDP is the least favorable probability of $\psi$ over all given strategies $\sigma$ of the MDP, i.e., $\min_\sigma \mathit{Pr}^{\sigma}(\psi)$ for a lower bound ($\geq$) and $\max_\sigma \mathit{Pr}^{\sigma}(\psi)$ for an upper bound ($\leq$).
If the worst case meets the threshold, then every given strategy of the MDP satisfies the requirement~\cite{Baier2008-principles}.

A \textit{permissive strategy} allows a non-empty subset of the available actions $A(s)$ in each state $s$ and leaves the choice among them open~\cite{Draeger2015-permissive}. {If the requirements' thresholds are met even with the worst-case choices of allowed actions, every strategy comprised of these actions satisfies the requirements.}

In the running example, $\mathit{at}_k$ labels the states where the agent is at $g_k$, and $\mathit{collision}$ those where it occupies the same cell as the moving obstacle.
With probability at least 0.8, the agent must reach each goal $g_k$, i.e., $P_{\geq 0.8}[\mathbf{F}\, \mathit{at}_k]$, and must not enter $g_k$ before reaching the earlier goals, e.g., $P_{\geq 0.8}[\neg \mathit{at}_2\ \mathbf{U}\ \mathit{at}_1]$.
While heading to $g_k$, it must also avoid the moving obstacle with probability at least 0.7, i.e., $P_{\geq 0.7}[\mathbf{G}\,(\mathit{seg}_k \Rightarrow \neg \mathit{collision})]$, where $\mathit{seg}_k$ holds in the states between reaching $g_{k-1}$ and reaching $g_k$.
\section{Problem Definition}
\label{sec:problem}
\label{sec:families}
\label{sec:rules}
\label{sec:certified}
\label{sec:adaptation-problem}
\textbf{Families and Their Vocabulary.}
We call a set $\mathcal{F}$ of instances of a parameterized MDP a \textit{family}, e.g., all grid worlds or those whose layouts follow one pattern, and write $M_I$ for the MDP of an instance $I \in \mathcal{F}$ (\autoref{sec:bg-mdp}).
The instances of $\mathcal{F}$ share a set of \textit{policy actions} $A_\pi$, a set of \textit{observable variables} and a set $\Phi$ of requirements.
We write $\theta_\varphi^I$ for the threshold $\theta$ of a requirement $\varphi \in \Phi$ on $I$, which may differ between instances.

{On an instance $I$, a \textit{controller} is a strategy of $M_I$ that selects a policy action in each state $s$ based only on the values $\mathit{obs}(s)$ of the observable variables.}
In the running example, the policy actions are the four moves, and the agent's cell, the goal flags and the moving obstacle's position are observable.

Values that differ between instances are not part of the state, so $\mathcal{F}$ also includes 
a vocabulary containing \emph{constants}, \emph{state features}, and \emph{action features}. 
\textit{Constants} are named values that each instance fixes, e.g., the row and column of each goal.
\textit{State features} are expressions over the observable variables and the constants, e.g., the number of moves left in a Pac-Man game.
\textit{Action features} are evaluated for each policy action. In the running example, \texttt{closer} holds for a move if it brings the agent closer to the next goal along a shortest path, \texttt{patrol} for a move into a cell on the moving obstacle’s patrol path, \texttt{wall} for a move into a wall or off the grid, \texttt{risky} for a move that can slip onto the patrol path or into a later goal, \texttt{obstacle} for a move onto the moving obstacle’s cell, and \texttt{early} for a move into a goal out of turn.

\textbf{Rule Lists.}
A \textit{generalized rule list} $R = \langle (c_1, a_1), \ldots, (c_n, a_n) \rangle$, or \textit{rule list} for short, is an ordered list of rules, each with a condition $c_j$ and an action $a_j \in A_\pi \cup \{\texttt{any}\}$.
Conditions are Boolean combinations of comparisons over the observable variables and the vocabulary.

On an instance $I$, a rule $(c_j, a_j)$ \textit{allows} a policy action $b$ in a state $s$ if $a_j \in \{b, \texttt{any}\}$ and $c_j$ holds in $\mathit{obs}(s)$ with every action feature in $c_j$ evaluated for $b$.
For example, rule~2 of \autoref{fig:intro-rules} allows every move that does not hit a wall, brings the agent closer to the next goal, cannot slip onto the patrol path or into a later goal, and neither meets the moving obstacle nor enters a goal out of turn.
A rule \textit{decides} a state if it is the first rule of $R$ that allows some policy action in that state.
{If the rule list $R$ does not have any rules that decide a given state, every possible action is allowed for it. } 
On each instance $I$, $R$ acts as a permissive strategy of $M_I$ (\autoref{sec:bg-pmc}), and we write $M_I^R$ for $M_I$ restricted to the actions that $R$ allows.

\textbf{Verified Rule Lists.}
We write $w_\varphi^I(R)$ for the worst case of a requirement $\varphi \in \Phi$ over all strategies of $M_I^R$.
A rule list $R$ is \textit{verified} on $I$ if $w_\varphi^I(R)$ meets the threshold $\theta_\varphi^I$ for every requirement $\varphi \in \Phi$.
If $R$ is verified on $I$, then every controller that takes only actions that $R$ allows satisfies all requirements in $\Phi$ on $I$ jointly, although each is checked separately.
This holds even if the controller uses memory or randomization.
At run time, the manager verifies and deploys each rule list in its \textit{frozen} form, which allows only the first of the actions that a rule allows in a fixed order of preference.

The \textit{margin} $m_\varphi^I(R)$ of $R$ on $I$ for $\varphi$ is $w_\varphi^I(R) - \theta_\varphi^I$ if $\theta_\varphi^I$ is a lower bound and $\theta_\varphi^I - w_\varphi^I(R)$ if it is an upper bound, i.e., how far the worst case lies on the safe side of the threshold.
The \textit{smallest margin} of $R$ on $I$ is $m^I(R) = \min_{\varphi \in \Phi} m_\varphi^I(R)$.

\textbf{Adaptation Problem.}
The system operates in \textit{episodes}, each of which is one run {of the system on} an instance of $\mathcal{F}$, e.g., the agent visiting the goals on one map.
\textbf{Given the parameterized MDP of a family $\mathcal{F}$, its requirements $\Phi$ and a vocabulary, the problem is to deploy, for each episode $t = 1, 2, \ldots$, a rule list verified on its instance $I_t \in \mathcal{F}$, where $I_t$ becomes known only just before episode $t$ starts. Since each episode waits for this decision, the time spent on it should be low.}
\section{Approach}
\label{sec:approach}
\begin{figure*}[t]
\centering
\resizebox{\textwidth}{!}{%
\begin{tikzpicture}[
  font=\small,
  >={Stealth[length=2mm]},
  proc/.style={draw, thick, fill=orange!18, minimum width=2.6cm, minimum height=1.1cm, align=center},
  art/.style={draw, rounded corners=3pt, fill=blue!8, minimum width=2.6cm, minimum height=0.75cm, align=center},
  dec/.style={draw, thick, rounded corners=8pt, fill=green!12, minimum width=2.7cm, minimum height=1.1cm, align=center},
  flow/.style={->, thick},
  lbl/.style={font=\fontsize{8.3}{9.6}\selectfont, align=center},
  mape/.style={font=\fontsize{8.3}{9.6}\selectfont\bfseries\itshape, text=black!70},
  region/.style={draw, dashed, rounded corners=6pt, inner sep=7pt},
  manager/.style={draw, rounded corners=8pt, inner sep=9pt}
]
\node[art, minimum width=2.4cm] (inst) at (-0.1, 0) {Instance $I_t$ of\\ the next episode};
\node[mape, above=3pt of inst] {Monitor};
\node[dec] (cur) at (3.9, 0) {Current $R_{\mathit{cur}}$\\ verified on $I_t$?};
\node[art] (lib) at (7.9, 1.55) {Library $\mathcal{L}$};
\node[mape, left=4pt of lib] {Knowledge};
\node[dec] (alt) at (7.9, 0) {Some $R \in \mathcal{L}$\\ verified on $I_t$?};
\node[mape, anchor=north] at (3.9, -0.62) {Analyze};
\node[art, minimum width=2.8cm] (dep) at (18.0, 0) {Rule list controlling\\ the episode on $I_t$};
\node[mape, above=3pt of dep, xshift=-0.55cm] {Execute};

\draw[flow] (inst) -- node[lbl, above] {model\\ $M_{I_t}$} (cur);
\draw[flow] (cur) -- node[lbl, above] {no} (alt);
\draw[flow] (lib) -- node[lbl, right] {stored rule lists} (alt);
\draw[flow] (cur.north) |- node[lbl, pos=0.62, above] {yes: retain $R_{\mathit{cur}}$} (18.6, 2.55) -- (18.6, 0.375);
\draw[flow] (alt) -- node[lbl, above] {yes: switch to the one with\\ the largest minimum margin} (dep);

\node[proc] (gen) at (4.1, -2.45) {B: Prompting\\ (LLM)};
\node[proc] (ver) at (8.3, -2.45) {C: Verification\\ (PRISM)};
\node[proc] (fb)  at (6.2, -4.25) {D: Feedback};
\node[proc] (sel) at (12.5, -2.45) {E: Selection\\ and repair};
\node[lbl, anchor=north east] (chains) at (13.85, -3.85) {repeated in $K$\\ independent chains};
\node[dec] (new) at (16.0, -2.45) {$R^\star$ verified\\ on $I_t$?};
\node[art] (fail) at (18.2, -4.3) {Adaptation fails,\\ episode deferred};

\draw[flow] (alt.south) -- ++(0, -0.85) -| node[lbl, pos=0.25, above] {no: generate on $T \ni I_t$} (gen.north);
\draw[flow] (gen) -- node[lbl, above] {rule\\ list $R$} (ver);
\draw[flow] (ver.south) |- node[lbl, pos=0.3, right] {worst-case values,\\ failure locations} (fb.east);
\draw[flow] (fb.west) -| node[lbl, pos=0.7, left] {revise\\ prompt} (gen.south);
\draw[flow] (ver) -- node[lbl, above] {candidates,\\ all rounds} (sel);
\draw[flow] (sel) -- node[lbl, above] {$R^\star$} (new);
\draw[flow] (new.east) -| node[lbl, pos=0.75, right] {yes: deploy,\\ add to $\mathcal{L}$} (dep.south);
\draw[flow] (new.south) |- node[lbl, pos=0.3, left] {no} (fail.west);

\begin{scope}[on background layer]
\node[region, fit=(gen)(ver)(fb)(sel)(chains), label={[mape, anchor=south east]north east:Plan: generation}] (genbox) {};
\node[manager, fit=(cur)(alt)(lib)(new)(genbox), label={[lbl, anchor=north east, xshift=-4pt, yshift=-3pt]north east:\textbf{A: Adaptation manager}}] {};
\end{scope}

\node[art, minimum width=1.1cm, minimum height=0.4cm] (la) at (-0.7, -2.8) {};
\node[lbl, anchor=west] at (la.east) {artifact};
\node[proc, minimum width=1.1cm, minimum height=0.4cm] (lp) at (-0.7, -3.5) {};
\node[lbl, anchor=west] at (lp.east) {process};
\node[dec, minimum width=1.1cm, minimum height=0.4cm, rounded corners=4pt] (ld) at (-0.7, -4.2) {};
\node[lbl, anchor=west] at (ld.east) {decision};
\end{tikzpicture}%
}
\caption{The adaptation manager, which picks the rule list for the episode on $I_t$ and invokes generation (dashed region) if no stored rule list is verified on $I_t$.}
\label{fig:overview}
\vspace{-0.15in}
\end{figure*}
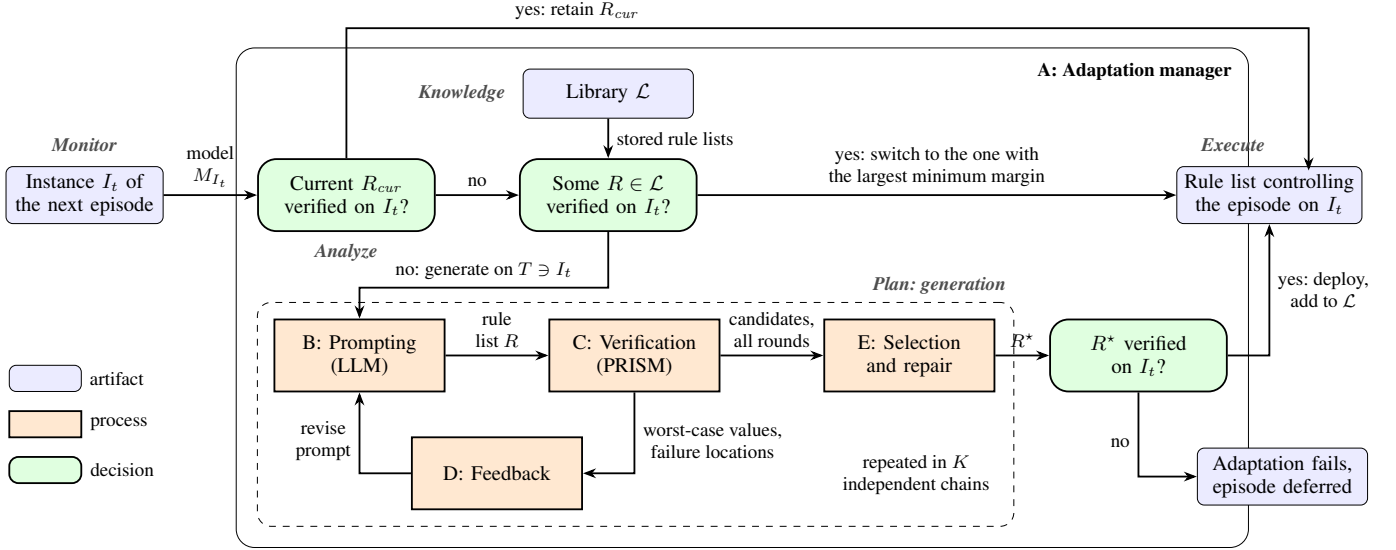

\autoref{fig:overview} illustrates the process of rule-list generation and application using model checkers.
Decisions are highlighted in \textcolor{green}{green}, processes in \textcolor{orange}{orange}, and artifacts, i.e., inputs, outputs and intermediate results, in \textcolor{blue}{blue}.
The \textbf{adaptation manager} (\autoref{sec:manager}) first checks whether its current rule list is verified on the next instance and, if so, retains it.
Otherwise, it switches to a verified rule list from its library.
If there is none, it generates a new rule list using a feedback-driven LLM rule list creation loop. First, an LLM writes a rule list (``\textbf{prompting}'', \autoref{sec:generation}), PRISM checks it on a few training instances, including the new one (``\textbf{verification}'', \autoref{sec:verification}), and the LLM is told where and why the rule list fails (``\textbf{feedback}'', \autoref{sec:feedback}).
``\textbf{Selection and repair}'' then improves the best rule lists from several such loops via small edits and chooses one (\autoref{sec:selection}).
If the result is verified on the instance, the manager deploys it and adds it to the library; otherwise, it defers the episode and records the failure or, optionally, falls back to per-instance synthesis.  
Before the first episode, {this generation procedure} automatically seeds the library from training sets of instances.

\subsection{Adaptation Manager}
\label{sec:manager}
\label{sec:certification}

\textbf{Run-Time Verification.}
Before the episode on $I_t$, the manager builds the instance's model $M_{I_t}$ from the observed parameters.
For a rule list $R$, it obtains the restricted model $M_{I_t}^R$ (\autoref{sec:rules}) by composing $M_{I_t}$ with one PRISM module that blocks every action that $R$ does not allow.
PRISM~\cite{Kwiatkowska2011-PRISM} then computes, by interval iteration~\cite{Haddad2018-interval-iteration}, a lower and an upper bound on the worst case $w_\varphi^{I_t}(R)$ of every requirement $\varphi \in \Phi$ on $M_{I_t}^R${, with sound value iteration~\cite{Quatmann2018-sound-VI} as a fallback upon an inconclusive result}.
The manager accepts $R$ as verified on $I_t$ only if, for every requirement $\varphi \in \Phi$, the less favorable of its two bounds meets $\theta_\varphi^{I_t}$.
Since each worst case lies between its two bounds, every accepted rule list is verified on $I_t$ (\autoref{sec:certified}).

\textbf{Adaptation Policy.}
Before each episode, the manager first checks whether its current rule list $R_{\mathit{cur}}$ is verified on $I_t$ and, if so, retains it, which avoids switching and costs a single verification.
Otherwise, it checks the other rule lists in its library $\mathcal{L}$ on $I_t$.
{The manager picks the rule list $R^{+}$ that comes closest to being verified, i.e., whose smallest margin over the requirements is the largest over the rule lists.}
If $R^{+}$ is verified on $I_t$, the manager switches to it, and otherwise it invokes generation.
It generates a rule list rather than synthesizing a strategy for $I_t$ alone, so that an instance that the library cannot serve extends the library.
The training set $T$ consists of $I_t$ and the three smallest training instances of $R^{+}$, so that the new rule list must also meet the requirements on instances for which $R^{+}$ was written.

\textbf{Unsuccessful Adaptation.}
Since generation is bounded, it can fail to produce a rule list that is verified on $I_t$ even if one exists.
The manager deploys the generated rule list $R^\star$ and adds it to $\mathcal{L}$ only if it is verified on $I_t$.
Otherwise, adaptation fails. By default, the manager then defers the episode on It and records the failure. Alternatively, it can fall back to synthesizing a strategy for $I_t$ alone with PRISM, which meets the requirements on $I_t$ whenever such a strategy exists. Since this strategy would only apply to $I_t$, the manager does not add it to $L$. In practice, PRISM exports such strategies only for queries with a single objective, so this fallback applies to Pac-Man and MER but not to the grid worlds, whose requirements must hold jointly.

\textbf{Running Example.}
{Suppose that the rule list of \autoref{fig:intro-rules} is not verified on the map of the next episode, since its worst case of reaching the second goal before the third is 0.793, just short of the threshold of 0.8, and that no other stored rule list is verified on this map either.
Since the manager can neither retain the current rule list nor switch to a stored one, it invokes generation on this map and the three training instances of the current rule list.
Generation returns a rule list that is verified with a worst case of 0.822 for the previously failing requirement, so the manager deploys it and adds it to $\mathcal{L}$.
It becomes the current rule list for the next episode.}

\subsection{Prompting}
\label{sec:generation}

The LLM receives a prompt that states the problem and asks for one rule list that meets the requirements, unchanged, on every training instance.
The prompt describes the family's dynamics, each training instance with its constants, a textual map (if it is provided by the domain), the vocabulary, the policy actions, the requirements with their thresholds, and how a rule list is read.

Since the rule list {may} be used unchanged on instances that are not shown to the LLM, including larger ones, the prompt asks for general rules and allows no number other than 0 and 1 in conditions {to avoid hard-coding values of a single instance.}
The LLM returns its answer as JSON constrained by a schema, which limits actions to the policy actions and \texttt{any}.
An answer whose rules do not parse or contain other numbers is rejected, and the LLM is asked again.

\subsection{Verification}
\label{sec:verification}

For each training instance $I \in T$, PRISM computes the worst case $w_\varphi^I(R)$ of every requirement $\varphi \in \Phi$ on $M_I^R$ (\autoref{sec:certification}).
A rule list $R$ \textit{passes} on a training instance $I \in T$ if the estimate of $w_\varphi^I(R)$ meets $\theta_\varphi^I$ for every requirement $\varphi \in \Phi$.

\subsection{Feedback}
\label{sec:feedback}

Prompting, verification and feedback form a loop;  we call one run of this loop a \textit{chain} and each of its iterations a \textit{round}. The chain stops once the rule list passes on every training instance with a smallest margin above a small constant, or after a fixed number of rounds.
Otherwise, the LLM receives a feedback prompt with the worst-case results on each training instance.
A state is \textit{lost} for a requirement if the outcome it asks for has probability 0 from that state under every strategy, e.g., after a collision.

{The feedback covers up to three failing requirements, choosing the ones whose worst cases fall furthest short of their thresholds.}
{For each of them, it picks the two states from which a run is most likely to step into a lost state.}  
For each such state, the feedback includes the deciding rule, the actions that rule allowed, and {the action leading to the worst case}. For every action, it also gives the features that hold for it and its risk of reaching a lost state in one step.
The prompt also lists the chain's earlier rule lists with their results, and the LLM returns a complete revised rule list.
The feedback never shows which action is optimal so that the LLM does not hard-code single decisions.

\begin{figure}[t]
\centering
\fbox{\begin{minipage}{\dimexpr\columnwidth-2\fboxsep-2\fboxrule\relax}
\footnotesize\ttfamily\raggedright
- avoid\_moving\_seg1 = 0.187 (needs >= 0.7).\\
\hspace*{1em}* 0.193 of runs fail from here: agent at (0,1), g1=false, g2=false, g3=false, obs\_idx=3 (obstacle at (1,2), next at (0,2)); decided by rule 3 `!g1 \& !wall \& !obstacle \& !early \& closer -> any'; allowed down; the worst case takes down.\\
\hspace*{2em}actions here: up: [wall, risky] risk: collision 0.15; right: [toward, closer, patrol, obstacle] risk: collision 0.70; down: [risky, toward, closer] risk: collision 0.15; left: [no feature true]
\end{minipage}}
\caption{An excerpt of the feedback on an earlier rule list while generating the rule list of \autoref{fig:intro-rules}, for the requirement to avoid the moving obstacle until the first goal. \texttt{toward} holds for a move that reduces the row or column distance to the next goal, ignoring walls and obstacles.}
\label{fig:feedback}
\vspace{-0.15in}
\end{figure}

In the state that \autoref{fig:feedback} shows for the smallest training instance, the deciding rule~3 allows only \texttt{down}, a move whose slips can collide with the moving obstacle, as its \texttt{risky} feature and its one-step risk show.
In its next answer, the LLM added \texttt{!risky} to that rule, which removes such moves wherever the rule applies on any instance.

\subsection{Selection and Repair}
\label{sec:selection}

A single chain may stall on a poor first answer, so once all $K$ chains end, we rank the distinct rule lists from all their rounds by their smallest margin over the training instances, $\min_{I \in T} m^I(R)$, and keep the highest-ranked ones as candidates.
Each candidate is then \textit{repaired} by hill climbing over small edits, such as dropping a rule, adding a conjunct, changing an action, or swapping two neighboring rules.
In each step, repair keeps the first edit that increases the sum of the margins $m_\varphi^I(R)$ over all $I \in T$ and $\varphi \in \Phi$.
{Repair computes these margins for many candidate rule lists in every generation run.}
{Thus, instead of calling PRISM for each candidate, we export the unrestricted model $M_I$ of each training instance from PRISM once and check each candidate $R$ in memory.}
{This in-memory checker obtains $M_I^R$ by removing the choices that $R$ disallows and computes the worst case of each requirement on it, without the cost of starting PRISM and building the model on every call.}
{Since it computes in floating point without an error bound, we use it only to rank candidates, and the manager verifies every rule list with PRISM before deploying it.}
Among the repaired candidates, we select the rule list that passes on the most training instances and, of those, {the one that is the furthest from failing any requirement.} 

{When the selected rule list is permissive, i.e., allows several actions, the deployed controller must choose one. A rule list that allows only the first of the allowed actions, given a fixed order of preference, is \emph{frozen}.}
{Selection freezes the rule list by using the same approximate checker as repair to rank action orders.}
Selection can also \textit{narrow} the frozen rule list by finding actions that fall short of an optimal one and greedily forbidding them, one place at a time (i.e., for one value of a variable whose range is the same on every instance, such as a maze cell), as long as the list still passes on every training instance.
\section{Evaluation}
\label{sec:evaluation}

We address the following research questions:  
\begin{enumerate}
  \item[\textbf{RQ1}] How effective are LLM-written rule lists on new instances, compared with learned controllers?
  \item[\textbf{RQ2}] How does the cost of our adaptation compare to per-instance solution synthesis?
  \item[\textbf{RQ3}] How are the adaptation manager’s decisions distributed among the available options across episode sequences?
  \item[\textbf{RQ4}] What is the effect of the feedback, repair, and narrowing components on our approach?
\end{enumerate}

\subsection{Evaluation Setup}
\label{sec:eval-setup}

\textbf{Domains.}
\textit{Grid-world navigation} (see \autoref{sec:bg-mdp}) stands in for a robot that visits goals in order while avoiding a moving obstacle, in buildings with different maps. \textit{Random grids} are maps whose obstacles are placed at random, and \textit{grid-world families} are three families of maps whose members each follow a layout pattern.
In the \textit{rooms} family, the goals lie in rooms joined by doors.
In the \textit{crossing} family, consecutive goals lie on opposite sides of a wall, and the moving obstacle patrols the wall's one gap.
In the \textit{adjacent} family, the first and third goals are neighbors.
In \textit{Pac-Man}~\cite{Hartmanns2019-QVBS}, the controller picks Pac-Man's direction at the crossings of a maze. Two ghosts move at random, and the only requirement is to keep the probability of being caught by a ghost at or below 0.56.
Its instances differ in the length of the game. We enable narrowing only on Pac-Man, whose training games are shorter versions of its test games.
{\textit{MER}~\cite{Feng2011-MER-model} models a Mars rover's resource arbiter, which shares communication and driving between two users and must never grant both at once.}
{As in 1--2--3--Go!~\cite{Azeem2025-123Go}, every move of a user and every answer of the arbiter is a policy action, so the controller is a scheduler that tries to make the arbiter fail by granting both resources at once.}
{The requirement for an instance $I$ is that the rule list makes the arbiter grant both resources at once with probability at least $0.95\,p^*_I$, where $p^*_I$ is the highest possible probability of granting both resources simultaneously on $I$.}
MER forms two benchmarks{, whose instances differ in the probability $x$ that a user keeps a resource and in the number of requests $N$.}

\textbf{Instances.}
Controllers are written, learned or trained on small \textit{training instances} and checked on disjoint \textit{held-out instances}, which are split into \textit{validation instances}, observed while we developed the prompts and the loop, and \textit{test instances}, on which we report our main results.
On random grids, controllers are trained on four sets of three grids of $9 \times 9$ to $11 \times 11$ and tested on 42 grids of up to $20 \times 20$, and on each family, they are trained on members of $6 \times 6$ to $8 \times 8$ and tested on 24 members of $15 \times 15$ to $20 \times 20$.
On Pac-Man, they are trained on games of length 14 to 18 and tested on lengths 50 and 100, and on each MER benchmark, they are trained on three small instances and tested on three further ones, with up to 100 requests.
The validation instances are eight random grids of $12 \times 12$ and $13 \times 13$, 20 members per family, six Pac-Man lengths from 8 to 30 and six instances per MER benchmark.
The thresholds of the grid worlds are the same at every size.
Every held-out instance admits a strategy that meets all requirements jointly, except possibly Pac-Man's length 100, whose unrestricted model $M_I$ could not be built.
On the grid worlds and MER, some controller we check is verified on each held-out instance, and at Pac-Man's validation lengths and length 50, an optimal strategy meets the threshold.

\textbf{Compared Controllers.}
LLM-written rule lists (\textbf{Full}) are generated as in \autoref{sec:approach} {by three LLMs, Claude Haiku~5.5, GPT-6 Luna and Gemini 3.8 Flash}\footnote{\texttt{anthropic/claude-haiku-5.5}, \texttt{openai/gpt-6-luna} and \texttt{google/gemini-3.8-flash} through OpenRouter, with reasoning and temperature set to the providers' defaults.}{, with the same loop and configuration}, in three runs per training set.
We report the frozen rule lists, which the manager deploys (\autoref{sec:selection}).
Unless stated otherwise, \textbf{Full} refers to the rule lists of Claude Haiku~5.5.
Our first baseline, \textbf{DT-raw}, {consists of} decision trees over raw state variables, learned from the optimal strategies of the training instances by our reimplementation of 1--2--3--Go!~\cite{Azeem2025-123Go}.
{Our second}, \textbf{DT-voc}, {consists of} decision trees over the vocabulary's features, learned in the same way, to separate the benefit of the vocabulary from that of LLM generation.
Our third baseline, \textbf{RL}, is a reinforcement learning policy trained over the vocabulary's features by proximal policy optimization (PPO)~\cite{Schulman2017-PPO}, as in work on generalizing across levels~\cite{Cobbe2020-Procgen}.
No baseline is learned or trained under a time limit. Trees and RL policies are checked as rule lists.

\textbf{Metrics.}
The \textit{verified coverage} of a controller is the fraction of a benchmark's test or validation instances on which it is verified.
We report its mean and standard deviation over the controllers of an approach. On the grid worlds, we also compute 95\% bootstrap confidence intervals of the difference in mean verified coverage between \textbf{Full} and \textbf{RL}, \textbf{DT-voc}, \textbf{Restart} and \textbf{NoRepair}.
We also report the manager’s decisions and their wall-clock time, the verified coverage of each rule list on its own training instances, and the mean number of LLM calls per training set.

\textbf{Implementation Details.}
Verification uses PRISM~4.10.1~\cite{Kwiatkowska2011-PRISM} with its explicit engine, interval iteration at precision $10^{-9}$ and a timeout of 900\,s, after which a check counts as not verified.
Where interval iteration does not converge or its bounds enclose a threshold, Storm~\cite{Hensel2022-Storm} decides the check by sound value iteration or, where that is inconclusive too, in exact rational arithmetic.
Generation runs $K = 6$ chains on the families and $K = 8$ elsewhere, each of up to seven rounds.
All experiment data and remaining settings, including prompts, generated rule lists and analysis scripts, are publicly available online~\cite{replication-package}.

\subsection{RQ1: Verified Coverage}
\label{sec:rq1}

{RQ1 measures the effectiveness of LLM-written rule lists (\textbf{Full}) on larger held-out instances as their verified coverage on every benchmark, compared with that of the learned baselines \textbf{DT-raw}, \textbf{DT-voc} and \textbf{RL}.}
{On the grid worlds, the features of \textbf{DT-voc} and \textbf{RL} include the goal flags so that every compared method observes the same information.}

\begin{table}[t]
\caption{Verified coverage (\%) of each approach on the test and validation instances, as mean $\pm$ standard deviation over its controllers. On the families, four of \textbf{RL}'s test checks timed out and count as not verified. Entries without a standard deviation come from a single controller. Best nonzero value per benchmark and column in bold.}
\label{tab:rq1-coverage}
\centering
\footnotesize
\begin{tabular}{@{}lrr@{}}
\toprule
Controller & Test & Validation \\
\midrule
\multicolumn{3}{@{}l}{\textit{Random grids}} \\
\multicolumn{3}{@{}l}{\textbf{Full}} \\
\quad Claude Haiku 5.5 & $\mathbf{91.3 \pm 7.3}$ & $\mathbf{92.7 \pm 11.3}$ \\
\quad GPT-6 Luna & $84.9 \pm 14.9$ & $86.5 \pm 18.0$ \\
\quad Gemini 3.8 Flash & $85.1 \pm 21.0$ & $91.7 \pm 18.7$ \\
\textbf{DT-raw} & $1.2 \pm 1.4$ & $0.0 \pm 0.0$ \\
\textbf{DT-voc} & $57.7 \pm 25.7$ & $46.9 \pm 27.7$ \\
\textbf{RL} & $77.0 \pm 12.7$ & $70.8 \pm 17.1$ \\
\midrule
\multicolumn{3}{@{}l}{\textit{Grid-world families}} \\
\multicolumn{3}{@{}l}{\textbf{Full}} \\
\quad Claude Haiku 5.5 & $89.4 \pm 9.8$ & $87.2 \pm 12.0$ \\
\quad GPT-6 Luna & $68.1 \pm 30.5$ & $73.9 \pm 27.7$ \\
\quad Gemini 3.8 Flash & $\mathbf{94.4 \pm 4.7}$ & $\mathbf{92.2 \pm 10.3}$ \\
\textbf{DT-raw} & $0.0 \pm 0.0$ & $0.0 \pm 0.0$ \\
\textbf{DT-voc} & $59.7 \pm 52.1$ & $61.7 \pm 45.4$ \\
\textbf{RL} & $60.2 \pm 45.3$ & $61.1 \pm 46.4$ \\
\midrule
\multicolumn{3}{@{}l}{\textit{Pac-Man}} \\
\multicolumn{3}{@{}l}{\textbf{Full}} \\
\quad Claude Haiku 5.5 & $0.0 \pm 0.0$ & $\mathbf{66.7 \pm 0.0}$ \\
\quad GPT-6 Luna & $0.0 \pm 0.0$ & $61.1 \pm 9.6$ \\
\quad Gemini 3.8 Flash & $0.0 \pm 0.0$ & $\mathbf{66.7 \pm 0.0}$ \\
\textbf{DT-raw} & $0.0$ & $\mathbf{66.7}$ \\
\textbf{DT-voc} & $0.0$ & $33.3$ \\
\textbf{RL} & $0.0 \pm 0.0$ & $50.0 \pm 0.0$ \\
\midrule
\multicolumn{3}{@{}l}{\textit{MER across $x$}} \\
\textbf{Full}, all three LLMs & $\mathbf{100.0 \pm 0.0}$ & $\mathbf{100.0 \pm 0.0}$ \\
\textbf{DT-raw} & $\mathbf{100.0}$ & $\mathbf{100.0}$ \\
\textbf{DT-voc} & $33.3$ & $50.0$ \\
\textbf{RL} & $0.0 \pm 0.0$ & $0.0 \pm 0.0$ \\
\midrule
\multicolumn{3}{@{}l}{\textit{MER across $N$}} \\
\textbf{Full}, all three LLMs & $\mathbf{100.0 \pm 0.0}$ & $\mathbf{100.0 \pm 0.0}$ \\
\textbf{DT-raw} & $\mathbf{100.0}$ & $\mathbf{100.0}$ \\
\textbf{DT-voc} & $0.0$ & $0.0$ \\
\textbf{RL} & $0.0 \pm 0.0$ & $0.0 \pm 0.0$ \\
\bottomrule
\end{tabular}
\vspace{-0.15in}
\end{table}

\begin{figure}[t]
\centering
\includegraphics[width=0.9\columnwidth]{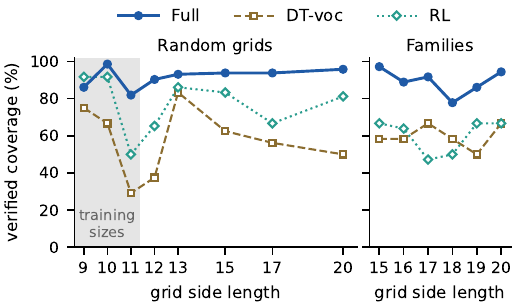}
\vspace{-0.1in}
\caption{Verified coverage on the test instances of the grid worlds by size, for \textbf{Full}, \textbf{DT-voc} and \textbf{RL}, as the mean over each approach's controllers, without spread. Random grids of $9 \times 9$ to $11 \times 11$ have the sizes of the training grids, and every other test instance is larger than any training instance. \textbf{DT-raw} is omitted.}
\vspace{-0.2in}
\label{fig:rq1-size}
\end{figure}

\begin{figure}[t]
\centering
\includegraphics[width=0.9\columnwidth]{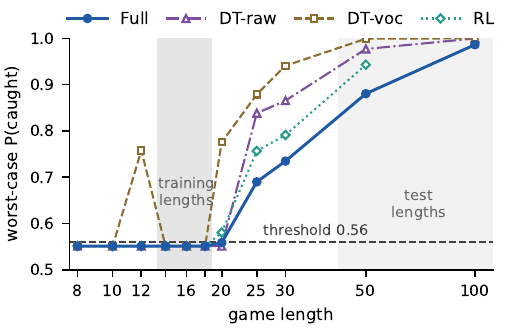}
\vspace{-0.1in}
\caption{Worst-case probability of being caught on Pac-Man by game length (logarithmic axis), for \textbf{Full} and \textbf{RL} (means of three runs or seeds; \textbf{RL} not checked at length 100), \textbf{DT-raw} and \textbf{DT-voc}, with the threshold of 0.56 dashed. Shading marks the training lengths 14 to 18 and the test lengths 50 and 100. Lines between measured lengths, e.g., from 30 to 50, are interpolated.}
\vspace{-0.15in}
\label{fig:rq1-pacman}
\end{figure}

\textbf{Results.}
\autoref{tab:rq1-coverage} gives the verified coverage of each controller on the test and validation instances of each benchmark. Our rule lists are small, with 17 to 46 rules on average per benchmark, whereas the trees of \textbf{DT-raw} and \textbf{DT-voc} have 256 to 36,031 nodes and the \textbf{RL} policies 5.4k to 7.8k parameters.

On both grid-world benchmarks, \textbf{Full} is verified on about 90\% of the test instances, more than every baseline. \textbf{Full}'s lead is statistically significant on both benchmarks, except over \textbf{DT-voc} on the families, whose coverage varies widely between families.
On the test instances, \textbf{Full}'s lead over \textbf{RL}, the strongest grid-world baseline, grows from 14.3 percentage points on random grids to 29.2 on the families.
On the families, \textbf{DT-voc} and \textbf{RL} each fail every test member of a family, whereas \textbf{Full} is verified on most members of every family.
\textbf{DT-raw} is verified on almost no held-out grid-world instance.
Its trees split on the raw cells of the training grids, so their decisions do not carry over to grids whose goals lie elsewhere.
\textbf{DT-voc}, which uses our vocabulary, reaches less than two thirds of \textbf{Full}'s coverage on the random test grids, meaning the vocabulary accounts for more than half of \textbf{Full}'s lead over \textbf{DT-raw}. Generation, i.e., the LLM together with verifier-guided repair, accounts for the rest.
\autoref{fig:rq1-size} shows the verified coverage on the grid worlds' test instances by size. 
On grid worlds larger than any training instance, \textbf{Full} is verified on at least 90\% of the random test grids and at least 78\% of the family members at every size, more than \textbf{DT-voc} and \textbf{RL} at each.

On MER, where the optimal actions hardly change across instances, \textbf{Full} and \textbf{DT-raw} are verified on every instance, whereas \textbf{RL} is verified on none, since its policies quickly reach a state whose best action is to maintain the current state.

\autoref{fig:rq1-pacman} shows the worst-case probability of being caught on Pac-Man by game length.
No controller is verified beyond length 20.
At length 50, the threshold lies just 0.009 above the optimum, and none of the compared controllers is near-optimal on games of length 25 and longer.
Beyond length 20, the rule lists of \textbf{Full} nonetheless come closest to the threshold among the controllers checked at each length, including both test lengths.

Rule lists written by GPT-6 Luna and Gemini 3.8 Flash are verified on 68\% to 94\% of the grid worlds' test instances, also more than every baseline on both benchmarks, and on every MER instance.
Gemini 3.8 Flash leads Claude Haiku~5.5 on the families and trails it on random grids.

\begin{tcolorbox}[boxsep=-1mm]
\textbf{RQ1.} Our LLM-written rule lists are verified on more held-out grid-world instances than every baseline, and they match the best baseline on MER. On Pac-Man's long test games, no controller is verified, but our rule lists come closest to the threshold.
\end{tcolorbox}

\subsection{RQ2: Cost of Adaptation vs. Synthesis}
\label{sec:rq2}
{RQ2 compares the time  of our adaptation with that of \textit{per-instance synthesis}. Per-instance synthesis computes a strategy that applies only to the instance at hand, whereas generation produces a rule list that the manager can reuse on other instances.
We measure the wall-clock time of PRISM's synthesis of a strategy on the unrestricted model $M_I$, of one verification of a stored rule list $R$ on $M_I^R$, and of generating one rule list. We use the validation sets of the random grids, the grid families, and Pac-Man lengths.
We derive the number of instances after which a rule list, counting its generation, costs less in total than synthesis per instance.}

\begin{figure}[t]
\centering
\includegraphics[width=0.9\columnwidth]{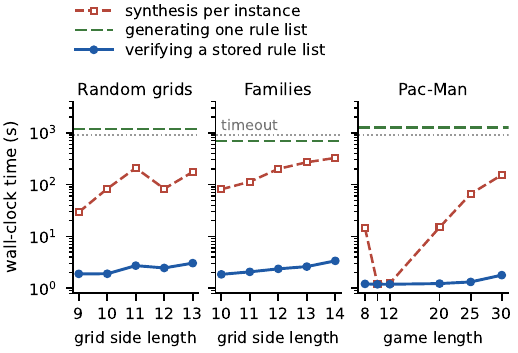}
\vspace{-0.1in}
\caption{Median wall-clock time of synthesizing a strategy for an instance and of verifying one stored rule list on it, by instance size, against the median time of generating one rule list. Synthesis is PRISM's multi-objective query for a strategy that meets all requirements jointly on the grid worlds, and its minimization of the probability of being caught on Pac-Man. It timed out at 900\,s on 4 of the 80 grid-world instances, which count as 900\,s. Generation runs without a timeout.}
\label{fig:rq2-synthesis}
\vspace{-0.2in}
\end{figure}

\textbf{Results.}
\autoref{fig:rq2-synthesis} gives the median times by instance size of per-instance synthesis and verifying a stored rule list, and of generating one rule list.
Retaining a rule list costs one verification, a median of 2.7\,s, and switching costs one verification per stored rule list.
Generating a new rule list takes a median of 808\,s, about 300 times as long as retaining one.
An episode on which no stored rule list is verified costs one verification per stored rule list and one generation.
This cost depends on the number of episodes only through the size of the library, which grows by one rule list per verified generation.
On the grid worlds, synthesis per instance takes 15 to 103 times as long as verifying a stored rule list (\autoref{fig:rq2-synthesis}), and the gap widens with size on the families.
These ratios are conservative under PRISM's synthesis timeout of 900\,s.
Synthesis takes longer at length 8 than at lengths 10 and 12 only because it was the first call of its run, which included about 11\,s of start-up time.
At length 100, the unrestricted model can no longer be built, whereas the restricted model of \textbf{Full}'s rule lists, with 14.1M states, can still be checked.
At every size we time, however, generating a rule list at run time takes longer at the median than synthesizing a strategy for the instance at hand, so generation pays off only through later reuse.

Counting the time to generate it, a rule list costs less in total than synthesis per instance once it serves several instances.
The larger the instances, the sooner this happens, since synthesis costs more as instance size increases.
For example, on the grid-world families, a rule list pays for its generation after about nine members of $10 \times 10$, but already after three of $14 \times 14$.

\begin{tcolorbox}[boxsep=-1mm]
\textbf{RQ2.} Verifying a stored rule list costs a small fraction of generating a new one or of synthesizing a strategy. Counting its generation, a rule list costs less than synthesis per instance once it serves several instances. 
\end{tcolorbox}

\subsection{RQ3: Adaptation Decision Distribution}
\label{sec:rq3}
{RQ3 measures what decisions the adaptation manager makes over sequences of episodes by counting how often it retains the current rule list, switches to another stored one, generates a new one, or defers the episode, given a starting set of seeded rule lists.}

\textbf{Sequences.}
We fix three sequences of 12 episodes,  $S_1$--$S_3$, that mix members of $10 \times 10$ to $14 \times 14$ with random grids of $9 \times 9$ to $13 \times 13$. $S_1$ runs four family members and then eight random grids, and $S_2$ and $S_3$ alternate several times between consecutive family members and consecutive random grids.
The sequences were chosen by hand before the runs.
To check that our findings do not depend on this choice, we also draw 10 sequences of 12 episodes and two longer sequences of 36 episodes, $L_1$ and $L_2$, at random from the pool of instances.
For each grid-world run, the manager starts from a new \textit{seed library} of three rule lists using \textbf{Full}, each generated from the training members of a single family.
No rule list in a seed library was generated from random grids, so the manager can serve a random-grid episode only by reusing a rule list written for a family or by generating a new one.
On Pac-Man, three sequences of 12 episodes run over the validation lengths, with the three rule lists of \textbf{Full} as the library. We run the manager without the synthesis fallback so that the deferred episodes show where generation fails.

\textbf{Results.}
\autoref{tab:rq3-sequences} counts the manager's decisions on each set of sequences for each seed library, in the \textbf{fixed} and \textbf{extended} settings.
Without generation, reuse, i.e., retaining or switching, serves about 85\% of the episodes of $S_1$--$S_3$ with each seed library and about 95\% of those of the random sequences, mostly by retaining the current rule list.
Where no stored rule list is verified, run-time generation serves most of the remaining grid-world episodes.
The rule lists it added were reused mostly on episodes on which a seed rule list was verified too. In $S_1$--$S_3$ and $L_2$, they served three episodes that would otherwise have needed a new generation.
On Pac-Man, the library serves every episode up to length 20, and the manager defers every episode of length 25 or 30, on which generation produces no verified rule list. With the synthesis fallback, the manager would serve these episodes instead. On the grid worlds, synthesis shows that a strategy that meets all requirements exists on every instance on which the manager defers an episode, so each deferral stems from generation.

\begin{tcolorbox}[boxsep=-1mm]
\textbf{RQ3.} The manager serves most episodes by reusing rule lists in the library, and run-time generation serves most of the remaining episodes. The rule lists that generation adds to the library occasionally serve an episode that no seeded rule list otherwise could.
\end{tcolorbox}

\begin{table}[t]
\caption{Decisions of the adaptation manager over sequences of episodes, by seed library (Lib.). In the \textbf{fixed} setting, generation is disabled. In the \textbf{extended} setting, the manager generates a rule list where no stored one is verified and adds it to the library once verified. Served counts the episodes on which the manager deploys a verified rule list, and Generated counts the verified generations out of those attempted. Episodes not served are deferred, since the synthesis fallback is disabled, and – marks a setting that was not run.}
\label{tab:rq3-sequences}
\centering
\footnotesize
\setlength{\tabcolsep}{3.5pt}
\begin{tabular}{@{}llrrrrr@{}}
\toprule
 & & \multicolumn{3}{c}{Fixed} & \multicolumn{2}{c}{Extended} \\
\cmidrule(lr){3-5}\cmidrule(l){6-7}
Sequences & Lib. & Retain & Switch & Served & Served & Generated \\
\midrule
$S_1$--$S_3$ (36) & 1 & 26 & 5 & 31 & 34 & 3/5 \\
 & 2 & 28 & 3 & 31 & 35 & 3/4 \\
 & 3 & 20 & 10 & 30 & 35 & 3/4 \\
\midrule
Random (120) & 1 & 102 & 12 & 114 & -- & -- \\
 & 2 & 104 & 11 & 115 & -- & -- \\
 & 3 & 78 & 35 & 113 & -- & -- \\
\midrule
$L_1$ (36) & 1 & 32 & 3 & 35 & 35 & 0/1 \\
$L_2$ (36) & 1 & 32 & 1 & 33 & 36 & 3/3 \\
\midrule
Pac-Man (36) & \textbf{Full} & 20 & 3 & 23 & 23 & 0/13 \\
\bottomrule
\end{tabular}
\vspace{-0.15in}
\end{table}

\subsection{RQ4: Feedback, Repair and Narrowing}
\label{sec:rq4}
{RQ4 evaluates the effect of the feedback, repair and narrowing components (\autoref{sec:approach}) on the verified coverage of our approach.}

\begin{table}[t]
\caption{Ablation of generation: verified coverage (\%) on the test, validation and training instances, as mean $\pm$ standard deviation over controllers, and mean LLM calls per training set, re-asks included. \textbf{Restart} was not run on MER across $N$.}
\label{tab:rq4-ablation}
\centering
\footnotesize
\setlength{\tabcolsep}{4pt}
\begin{tabular}{@{}lrrrr@{}}
\toprule
Configuration & Test & Validation & Training & Calls \\
\midrule
\multicolumn{5}{@{}l}{\textit{Random grids}} \\
\textbf{Full} & $91.3 \pm 7.3$ & $92.7 \pm 11.3$ & $91.7 \pm 15.1$ & 55.8 \\
\textbf{Restart} & $77.0 \pm 12.0$ & $79.2 \pm 9.7$ & $75.0 \pm 20.7$ & 56.0 \\
\textbf{NoRepair} & $85.7 \pm 5.9$ & $86.5 \pm 6.4$ & $80.6 \pm 17.2$ & 55.8 \\
\textbf{OneShot} & $76.2 \pm 4.7$ & $74.0 \pm 3.6$ & $50.0 \pm 17.4$ & 1.0 \\
\midrule
\multicolumn{5}{@{}l}{\textit{Grid-world families}} \\
\textbf{Full} & $89.4 \pm 9.8$ & $87.2 \pm 12.0$ & $100.0 \pm 0.0$ & 36.4 \\
\textbf{Restart} & $75.0 \pm 28.6$ & $71.7 \pm 26.9$ & $100.0 \pm 0.0$ & 42.0 \\
\textbf{NoRepair} & $71.8 \pm 36.0$ & $68.3 \pm 39.1$ & $88.9 \pm 16.7$ & 36.4 \\
\textbf{OneShot} & $43.1 \pm 43.1$ & $43.9 \pm 35.8$ & $37.0 \pm 30.9$ & 1.0 \\
\midrule
\multicolumn{5}{@{}l}{\textit{Pac-Man}} \\
\textbf{Full} & $0.0 \pm 0.0$ & $66.7 \pm 0.0$ & $100.0 \pm 0.0$ & 49.3 \\
\textbf{Restart} & $0.0 \pm 0.0$ & $33.3 \pm 28.9$ & $33.3 \pm 57.7$ & 59.0 \\
\textbf{NoRepair} & $0.0 \pm 0.0$ & $50.0 \pm 0.0$ & $100.0 \pm 0.0$ & 49.3 \\
\textbf{OneShot} & $0.0 \pm 0.0$ & $38.9 \pm 9.6$ & $33.3 \pm 57.7$ & 1.0 \\
\midrule
\multicolumn{5}{@{}l}{\textit{MER across $x$}} \\
\textbf{Full} & $100.0 \pm 0.0$ & $100.0 \pm 0.0$ & $100.0 \pm 0.0$ & 63.7 \\
\textbf{Restart} & $100.0 \pm 0.0$ & $100.0 \pm 0.0$ & $100.0 \pm 0.0$ & 59.0 \\
\textbf{NoRepair} & $100.0 \pm 0.0$ & $100.0 \pm 0.0$ & $100.0 \pm 0.0$ & 63.7 \\
\textbf{OneShot} & $0.0 \pm 0.0$ & $0.0 \pm 0.0$ & $0.0 \pm 0.0$ & 1.0 \\
\midrule
\multicolumn{5}{@{}l}{\textit{MER across $N$}} \\
\textbf{Full} & $100.0 \pm 0.0$ & $100.0 \pm 0.0$ & $100.0 \pm 0.0$ & 60.0 \\
\textbf{NoRepair} & $100.0 \pm 0.0$ & $100.0 \pm 0.0$ & $100.0 \pm 0.0$ & 60.0 \\
\textbf{OneShot} & $33.3 \pm 57.7$ & $33.3 \pm 57.7$ & $44.4 \pm 50.9$ & 1.0 \\
\bottomrule
\end{tabular}
\vspace{-0.15in}
\end{table}
\textbf{Ablation Configurations.}
\textbf{Restart} replaces feedback with independent restarts from the initial prompt, with the same limits on the number of iterations.
\textbf{NoRepair} skips repair and narrowing (\autoref{sec:selection}).
\textbf{OneShot} deploys the first rule list that the LLM writes without further processing.

\textbf{Results.}
To evaluate the effect of feedback, we compare our \textbf{Full} approach with \textbf{Restart} and \textbf{OneShot}, which do not use feedback.
\autoref{tab:rq4-ablation} compares \textbf{Full} with \textbf{OneShot} and \textbf{Restart} by verified coverage and by the mean number of LLM calls per training set.
We find that \textbf{OneShot} is verified on fewer instances than \textbf{Full} on every benchmark except Pac-Man's test lengths, and, on average, on at most half of its own training instances.
On random test grids, \textbf{Restart} barely improves on \textbf{OneShot}, so the gain there comes from feedback, whereas on the families, \textbf{Restart} closes most of the gap between \textbf{OneShot} and \textbf{Full}. \textbf{Full}'s lead over \textbf{Restart} is statistically significant on random grids.
The lead of feedback over independent restarts contrasts with deterministic tasks in which resampling the LLM was  as effective as feeding back errors at matched budgets~\cite{Stechly2025-self-verification,Olausson2024-self-repair}.

\autoref{tab:rq4-ablation} also gives the verified coverage of \textbf{NoRepair} alongside that of \textbf{Full}.
Without repair, test coverage drops by 5.6 percentage points on random grids and by 17.6 on the families. The drop on random grids is statistically significant. We believe that these drops occur because the rule list selected before repair often falls just short of a threshold on one training instance, a gap that a single edit, such as adding \texttt{!risky} to a rule, closes on every instance at once. On the families, feedback and repair raise mean coverage by at least as much as on random grids, but coverage varies more between runs.
Of the other two LLMs, repair raises test coverage only for GPT-6 Luna on the families.
On Pac-Man, repair changes no worst case at any length, and the gain over \textbf{NoRepair} comes entirely from narrowing, which lowers the worst case at length 20 from 0.662 to 0.559, just below the threshold.

\begin{tcolorbox}[boxsep=-1mm]
\textbf{RQ4.} Under the same cap of chains and rounds, verifier feedback raises verified coverage over independent restarts (\textbf{Restart}) by about 14 percentage points on the test instances of random grids. It also leads on Pac-Man's validation lengths and ties on MER across $x$. In contrast to reports on deterministic tasks~\cite{Stechly2025-self-verification,Olausson2024-self-repair}, feedback pays off here. Repair raises verified coverage on the test instances of random grids, though not with every LLM, and narrowing raises it on Pac-Man's validation lengths.
\end{tcolorbox}

\subsection{Threats to Validity}
\label{sec:threats}

\textbf{Internal Validity.}
Randomness in LLM outputs and in PPO training could affect the comparison between approaches.
We mitigate this by using three runs per training set, reporting standard deviations, and testing the differences on the grid worlds for statistical significance.
LLM providers may change or retire their models, which limits reproducibility.
We mitigate this by giving the model identifiers (\autoref{sec:eval-setup}) and by releasing every prompt and generated rule list~\cite{replication-package}.
We developed the prompts and the loop's parameters based on preliminary results on the validation instances, so the results that use no test instances, i.e., the sequences of RQ3, may be optimistic.
However, on the grid worlds and MER, the coverage of the validation instances differs from that on the test instances by at most 6.6 percentage points, which implies that this tuning inflates results on the validation instances only slightly.
Floating-point rounding could break the bounds of interval iteration on which verification relies~\cite{Hartmanns2022-floating-point}.
We mitigate this in part by deciding checks whose bounds enclose a threshold by sound value iteration or exact arithmetic.
Verification also depends on our translation of rule lists and of the vocabulary's features into PRISM.
Unit tests check this translation against the first-match semantics of \autoref{sec:rules} on small models.
As a second check, repair's checker, which evaluates rule lists without PRISM, agrees with PRISM's worst cases to within $10^{-6}$ in every domain. 

\textbf{External Validity.}
Our three domains are benchmarks that stand in for self-adaptive systems whose environments vary in size and structure, so our findings may not carry over to other systems or LLMs.
The LLMs may also have seen our domains' models during training, since all three have been published.
Memorized solutions cannot be reused directly, however, since none is written in our rule language or vocabulary, and every rule list is checked on each instance's model.
The benefit of the LLM may depend on the vocabulary, which we wrote by hand for each domain and did not vary.
\textbf{DT-voc} separates the vocabulary's contribution from that of generation, but not how generation would fare with another vocabulary.
The framework assumes exact models, so its guarantees hold only relative to these models.
Explicit-state model checking limits the instance sizes that we can verify.
For restricted rule lists, however, the model is much smaller, so the rule lists of \textbf{Full} could still be checked at Pac-Man's length 100, whose unrestricted model could not be built (\autoref{sec:rq2}).

\textbf{Construct Validity.}
We measure verified coverage rather than the outcome of episodes, because the manager deploys only verified rule lists.
Verified coverage depends on where the thresholds lie, since a controller that narrowly misses one counts as not verified.
We take the grid-world thresholds from prior work~\cite{Arya2026-gridworld-PMC} and set Pac-Man's just above its optimum.
MER's thresholds come from each instance's optimum, which a deployed manager would not know, and stand in for thresholds set by stakeholders.
Verified coverage also counts an instance on which no controller can meet the requirements as a miss, which may happen only at Pac-Man's length 100.
On Pac-Man, we therefore also compare the worst cases themselves, which do not depend on the threshold.
We report the times of verification, generation and synthesis separately, so that each can be rescaled to other hardware, and we ran all timed experiments on a dedicated machine.
\section{Related Work}
\label{sec:related}

\textbf{LLMs in Self-Adaptation with Verification.}
LLMs have been surveyed as aids across the MAPE-K loop~\cite{Li2024-LLMs-in-SAS,Li2024-GenAI-SAS-roadmap} and used to reason about unanticipated situations~\cite{Alves2026-MAPER}, to refine safety operational rules~\cite{Gaaloul2026-grammar-constrained-rules}, and to explain task plans~\cite{Vazquez2026-mind-the-prompt}.
Gross et al.\ verify the decisions of an LLM acting as the policy itself on a single model~\cite{Gross2025-verifying-LLM-decisions}, and use an LLM to repair the unsafe decisions of a learned policy in a single round~\cite{Gross2024-counterfactual-LLM}.
Other approaches use a model checker qualitatively~\cite{Lee2025-VeriPlan}, check a model learned from traces at run time~\cite{Wang2026-ProbGuard}, or let an LLM write the MDP and a model checker synthesize the strategy~\cite{Saccon2026-MDP-generation}.
Closest to our use of LLMs, T\"{o}pfer et al.~\cite{Topfer2026-vibe-coding} have LLMs write adaptation managers that are revised from constraints violated in test runs, and Arya et al.~\cite{Arya2026-gridworld-PMC} refined LLM-written per-state policies for single grid-world instances over several rounds of quantitative model-checking results.
In our framework, the LLM instead writes one rule list for a stochastic environment for reuse across the instances of a family, and the manager reuses it only where probabilistic model checking verifies it.

\textbf{Run-Time Verification and Adaptation.}
Quantitative verification at run time has long guided adaptation decisions~\cite{Calinescu2012-CACM-runtime-QV,Moreno2015-proactive,Jamshidi2019-ML-meets-planning}.
Frameworks such as Rainbow~\cite{Garlan2004-Rainbow} select at run time among adaptation strategies written in advance~\cite{Cheng2012-Stitch}, which probabilistic model checking can analyze~\cite{Camara2014-stochastic-games}.
Earlier plans~\cite{Kinneer2018-plan-reuse,Kinneer2021-information-reuse}, repertoires of plans built in advance~\cite{Kinneer2020-repertoires}, and models carried over to new environments by transfer~\cite{Jamshidi2017-transfer-learning} or lifelong learning~\cite{Gheibi2022-lifelong} seed a new search or learning step, whereas our manager reuses rule lists unchanged wherever they are verified, and otherwise generates a new one.

\textbf{LLM-Written Generalized Plans.}
Generalized plans are programs or policies that solve every instance of a planning domain.
Silver et al.~\cite{Silver2024-GenPlan-LLM} showed that LLMs can write such programs when debugged on training tasks.
Later work refined how LLMs write such programs~\cite{Stein2026-strategy-refinement,Murray2026-GenePlan} or checked a property of LLM-written heuristics~\cite{Pereira2026-property-guided}.
Closest to our loop is Prompt, Prove, Patch~\cite{Drexler2026-PromptProvePatch}, which proves that an LLM-written general policy reaches the goal on every training task and returns counterexamples to the LLM.
These generalized planners target deterministic domains, where running a policy on an instance verifies it.
They check that a policy generalizes to larger instances by running it on some of them or by hand-writing proofs~\cite{Stein2026-provably-complete}.
Our loop has the same generate--verify--revise shape, but judges a rule list on stochastic models by its worst case, and our manager verifies it automatically before reusing it on a new instance.

\textbf{Verifier Feedback and Resampling.}
LLM-Modulo~\cite{Kambhampati2024-LLM-Modulo} and other neuro-symbolic frameworks place an LLM inside a loop of sound critics.
In deterministic tasks, resampling the LLM was about as effective as feeding back errors~\cite{Stechly2025-self-verification,Olausson2024-self-repair}.
We compare verifier feedback with independent restarts on stochastic models, where the verifier's output is quantitative (\autoref{sec:rq4}).

\textbf{Generalized Policies.}
Closest to our rule lists, the general policies of \cite{Bonet2015-general-policies,Bonet2018-features} are rules over Boolean and numerical features, and their feature effects play the role of our action features such as \texttt{closer}.
Later work learns such features and policies from small examples~\cite{Frances2021-general-policies}, whereas our vocabulary is written by the domain's designer.
Relational decision lists~\cite{Yoon2002-inductive-policy,Fern2006-decision-lists} share the ordered form of our rule lists but are evaluated by simulation, without thresholds or verification.

\textbf{Families of MDPs and Permissive Strategies.}
Policy trees~\cite{Andriushchenko2024-PoliciesGrowOnTrees} and satisfiability modulo probabilistic model checking~\cite{Heck2026-SMPMC} synthesize strategies that meet a specification on every member of a finite family, and robust methods optimize a policy's worst case over a finite family of partially observable MDPs~\cite{Galesloot2025-rfPG} or bound the risk on unseen MDPs sampled from a distribution~\cite{Schnitzer2025-robust-parametric}.
The members of these families share one state space, whereas the instances of our families differ in size and structure.
1--2--3--Go!~\cite{Azeem2025-123Go} learns decision trees from the optimal strategies of small instances of a parameterized MDP and computes their value exactly on each larger instance, for a single reachability objective.
Its trees split on raw state variables, so they carry over only where the right decision does not depend on the instance's parameters (\autoref{sec:rq1}), whereas our vocabulary names instance values, and our rule lists are verified on each instance against several requirements.
Permissive strategies~\cite{Draeger2015-permissive,Junges2016-permissive} and shields~\cite{Alshiekh2018-shielding,Jansen2020-probabilistic-shields,Heck2026-probabilistic-shields} allow sets of actions in each state while preserving guarantees, and tools such as dtControl~\cite{Ashok2021-dtControl} compress a strategy into a small decision tree, in each case for a single model.
\section{Conclusion}
\label{sec:conclusion}
In this paper, we proposed a verifier-guided framework for adaptation through controller reuse and generation across families of stochastic environments.
In this framework, an LLM writes generalized rule lists over domain-level features and refines them with quantitative feedback from probabilistic model checking.
At run time, an adaptation manager reuses a rule list from its library only if model checking verifies it on the instance at hand, and otherwise generates a new one, which joins the library once verified.

We evaluated our framework with three LLMs on grid-world navigation, Pac-Man and MER.
On the grid worlds, the LLM-written rule lists were verified on more held-out instances than every baseline, even on instances larger than any training instance.
They matched the best baseline on MER and came the closest to the threshold on Pac-Man's long test games, where no controller was verified.
Over sequences of grid-world instances, reuse served most episodes at a small fraction of the cost of generation or of synthesis per instance.
Including the cost of generating it, a rule list became cheaper than synthesis once it had served several instances.
The rule lists that run-time generation added to the library occasionally served episodes that would otherwise have needed a new generation.
Under the same cap of chains and rounds, verifier feedback reached a higher verified coverage than independent restarts on random grids and on Pac-Man's validation lengths, and repairing rule lists by small edits raised it on random grids, though not with every LLM.

As future work, we plan to evaluate over longer sequences whether the rule lists that run-time generation adds to the library reduce the need for later generations.
We also aim to let the LLM write the vocabulary, which we wrote by hand for each domain, and to verify stored rule lists against models updated from run-time data, as SAVE does~\cite{Vazquez2026-SAVE}.
To speed up the verification of stored rule lists, we plan to use parametric~\cite{Filieri2011-runtime-PMC}, incremental~\cite{Kwiatkowska2011-incremental} and caching~\cite{Gerasimou2014-caching} techniques, which reduce the cost of quantitative verification at run time.
Finally, we plan to verify rule lists on instances too large for explicit-state model checking, e.g., with symbolic engines~\cite{Kwiatkowska2004-PRISM-hybrid}, or otherwise check them by statistical model checking~\cite{Agha2018-SMC-survey}.

\bibliographystyle{IEEEtran}
\bibliography{refs}

\end{document}